\documentclass[12pt]{JHEP3}

\usepackage{epsfig, psfig}

\def\spose#1{\hbox to 0pt{#1\hss}}
\def\ltapprox{\mathrel{\spose{\lower 3pt\hbox{$\mathchar"218$}}
 \raise 2.0pt\hbox{$\mathchar"13C$}}}
\def\gtapprox{\mathrel{\spose{\lower 3pt\hbox{$\mathchar"218$}}
 \raise 2.0pt\hbox{$\mathchar"13E$}}}

\title{
The 4D SU(3) gauge theory with an imaginary $\theta$ term
}

\author{Haralambos Panagopoulos \\ 
        Department of Physics, University of Cyprus,
        Lefkosia, CY-1678, Cyprus\\ 
        E-mail: \email{haris@ucy.ac.cy} 
}

\author{Ettore Vicari \\ 
        Dipartimento di Fisica, Universit\`a 
        di Pisa and INFN, I-56127 Pisa, Italy \\ 
        E-mail: \email{vicari@df.unipi.it} 
}

\abstract{

We study the scaling behavior of the 4D SU(3) lattice gauge theory in
the presence of a $\theta$ term, by Monte Carlo simulations
computing the topological properties at imaginary $\theta$.  The
numerical results provide a good evidence of scaling in the continuum
limit.  The imaginary $\theta$ dependence of the ground-state energy
turns out to be well described by the first few terms of related
expansions around $\theta=0$, providing accurate estimates of the
first few coefficients, up to $O(\theta^6)$.

}

\keywords{Gauge Field Theories, Lattice Gauge Field Theories, Theta
Term, Topological Term}

\begin{document}

Four-dimensional $SU(N)$ gauge theories have a nontrivial dependence
on the parameter $\theta$ which appears in the Euclidean Lagrangian as
\begin{equation}
{\cal L}_\theta  = {1\over 4} F_{\mu\nu}^a(x)F_{\mu\nu}^a(x)
- i \theta {g^2\over 64\pi^2} \epsilon_{\mu\nu\rho\sigma}
F_{\mu\nu}^a(x) F_{\rho\sigma}^a(x),
\label{lagrangian}
\end{equation}
where 
\begin{equation}
q(x)=\frac{g^2}{64\pi^2} \epsilon_{\mu\nu\rho\sigma} F_{\mu\nu}^a(x)
F_{\rho\sigma}^a(x)
\label{topchden}
\end{equation}
is the topological charge density.  The $\theta$ dependence of the
ground-state energy density $F(\theta)$, defined as
\begin{equation}
Z_\theta = 
\int [dA] \exp \left(  - \int d^4 x {\cal L}_\theta \right)
=\exp[ - V F(\theta) ],
\label{vftheta}
\end{equation}
where $V$ is the space-time volume,  can be written as
\begin{eqnarray}
{\cal F}(\theta)\equiv
F(\theta)-F(0)={1\over 2} \chi \theta^2 s(\theta),\label{ftheta}
\end{eqnarray}
where $\chi$ is the topological susceptibility at $\theta=0$, 
\begin{equation}
\chi = \int d^4 x \langle q(x)q(0) \rangle_{\theta=0} 
= {\langle Q^2 \rangle_{\theta=0} \over V},\quad
Q\equiv \int d^4x\, q(x),
\label{chidef}
\end{equation}
and $s(\theta)$ is a dimensionless even function of $\theta$ such that
$s(0)=1$.  Assuming analyticity at $\theta=0$, $s(\theta)$ can be
expanded as
\begin{eqnarray}
s(\theta) = 1 + b_2 \theta^2 + b_4 \theta^4 + \cdots,
\label{stheta}
\end{eqnarray}
where only even powers of $\theta$ appear.  Large-$N$ scaling
arguments~\cite{Hooft-74,Witten-79b,Witten-80} applied to the
Lagrangian (\ref{lagrangian}) of a 4D SU($N$) gauge theory indicate
that the relevant scaling variable in the large-$N$ limit is
$\bar\theta\equiv {\theta/N}$.  This implies that in the large-$N$
limit $\chi=O(1)$, while the coefficients $b_{2i}$ are suppressed by
powers of $N$, i.e.~\cite{VP-09} 
\begin{equation}
b_{2i}=O(N^{-2i}).
\label{b21}
\end{equation}

The presence of the $\theta$ term has important phenomenological
consequences, since it violates parity and time reversal symmetry.
Experimental bounds on the $\theta$ parameter in QCD are best obtained
from the electric dipole moment of the
neutron~\cite{Baker,Harris-etal-99,VP-09}, which leads to an
unnaturally small value for $\theta$, $|\theta| \ltapprox 10^{-10}$.
This suggests the idea that there must be a mechanism responsible for
suppressing the value of $\theta$ in the context of QCD.  However, the
issue of the $\theta$ dependence, even within pure gauge theory, has
phenomenological relevance because it provides an explanation to the
so-called U(1)$_A$ problem~\cite{Hooft-76,Witten-79,Veneziano-79},
i.e.  explaining the large mass of the $\eta'$ meson with respect to
the $\pi$ meson.  The $\theta$ dependence of 4D SU($N$) gauge theories
is particularly interesting in the large-$N$ limit where the issue may
also be addressed by other approaches, such as AdS/CFT correspondence
applied to nonsupersymmetric and non conformal theories, see
e.g. Refs.~\cite{AGMOO-00,Witten-98,DiVecchia-99,GK-07,GKN-07}.

Due to the nonperturbative nature of the $\theta$ dependence,
quantitative assessments of this issue have largely focused on the
lattice formulation of the theory, using Monte Carlo (MC) simulations.
However, the complex nature of the $\theta$ term in the Euclidean QCD
Lagrangian prohibits a direct MC simulation at $\theta\ne 0$.
Information on the $\theta$ dependence of physically relevant
quantities, such as the ground state energy and the spectrum, has been
obtained by computing the coefficients of the corresponding expansions
around $\theta = 0$.  The coefficients of expansion of $s(\theta)$,
cf. Eq.~(\ref{stheta}), can be determined from appropriate
zero-momentum correlation functions~\cite{VP-09,DPV-02} of the
topological charge density at $\theta=0$, which are related to the
moments of the $\theta=0$ probability distribution $P(Q)$ of the
topological charge $Q$.  Indeed
\begin{eqnarray}
&&b_2 = - {\chi_4 \over 12 \chi}, \qquad  \chi_4 = {1\over V} \left[ 
\langle Q^4 \rangle - 3  \langle Q^2 \rangle^2 \right]_{\theta=0}, 
\label{b2chi4} \\
&&b_4 = - {\chi_6\over 360 \chi},\qquad
\chi_6 = {1\over V} \left[ 
\langle Q^6 \rangle  -  15 \langle Q^2 \rangle \langle Q^4 \rangle  +
30 \langle Q^2 \rangle^3 \right]_{\theta=0},  \label{b4chi6} 
\end{eqnarray}
etc. They parameterize the deviations of $P(Q)$ from a simple Gaussian
behavior.  As shown in Ref.~\cite{Luscher-04}, the correlation
functions involving multiple zero-momentum insertions of the
topological charge density can be defined in a nonambiguous,
regularization independent way, and therefore the expansion
coefficients $b_{2i}$ are well defined renormalization group invariant
quantities.  The numerical evidence for a nontrivial
$\theta$-dependence, obtained through MC simulations of the lattice
formulation, appears quite robust.  We refer the reader to
Ref.~\cite{VP-09} for a recent review.  On the other hand, MC
simulations at $\theta=0$ have only made it possible to estimate the
ground-state energy up to the $O(\theta^4)$ term, because the
statistical errors rapidly increase with increasing order of the
expansion, essentially for importance sampling problems.  The
large-$N$ prediction $b_2=O(N^{-2})$ has been already supported by
numerical results~\cite{DPV-02,Delia-03,GPT-07}.  The calculation of
the higher-order terms would provide a further check of the large-$N$
arguments, which predict them to be suppressed by higher powers of
$N$, as in Eq.~(\ref{b21}).

In this paper we consider imaginary values of $\theta$, which make the
Euclidean Lagrangian (\ref{lagrangian}) real, thus allowing us to
investigate the imaginary $\theta$ dependence by MC simulations.
Assuming analyticity at $\theta=0$, the results may provide
quantitative information on the expansion around $\theta=0$,
cf. Eqs.~(\ref{ftheta}) and (\ref{stheta}).  Indeed, fits of the data
to polynomials of imaginary $\theta$ may provide more accurate
estimates of the coefficients, overcoming the rapid increasing of the
statistical errors observed at $\theta=0$.  Another interesting issue
concerns the continuum limit of the theory for generic, and in
particular imaginary, $\theta$ values.  Perturbative
renormalization-group (RG) arguments~\cite{ET-82} indicate that
$\theta$ is a RG invariant parameter of the theory, thus the continuum
limit should be approached while keeping $\theta$ fixed to any complex
value.  Although reflection positivity~\cite{MM-94}, and therefore a
well defined relativistic field theory, requires real $\theta$ values,
imaginary $\theta$ may still give rise to a well defined continuum
limit in the sense of a statistical field theory, with the same
asymptotic behavior controlled by the perturbative $\beta$ function
which does not depend on $\theta$.  As we shall see, this is indeed
supported by the numerical data for the 4D SU(3) lattice gauge theory,
at least for $|\theta|<\pi$, which are well described by the first few
nontrivial terms of the expansion around $\theta=0$.

Introducing the real parameter $\theta_i$, defined by
\begin{equation}
\theta \equiv - i \theta_i,
\label{thetaidef}
\end{equation} 
Eq.~(\ref{ftheta}) becomes
\begin{eqnarray}
\Phi(\theta_i) \equiv {\cal F}(-i\theta_i) = - {1\over 2} \chi
\theta_i^2 s(-i\theta_i) = - {1\over 2} \chi \theta_i^2 \left(1 - b_2
\theta_i^2 + b_4 \theta_i^4 + \cdots\right).
\label{Fthetai}
\end{eqnarray}
We thus obtain
\begin{eqnarray}
&& {\langle Q \rangle_{\theta_i}\over V} = -{\partial
\Phi(\theta_i)\over \partial \theta_i} = \chi \theta_i \left( 1 - 2
b_2 \theta_i^2 + 3 b_4 \theta_i^4 + \cdots\right),
\label{qthetai}\\
&& {\langle Q^2 \rangle^{c}_{\theta_i} \over V} \equiv {\langle Q^2
\rangle_{\theta_i} - \langle Q \rangle_{\theta_i}^2 \over V} =
-{\partial^2 \Phi(\theta_i)\over \partial \theta_i^2} = \chi \left( 1
- 6 b_2 \theta_i^2 + 15 b_4 \theta_i^4 + \cdots\right),
\label{chithetai}
\end{eqnarray}
etc.

The 4D SU($N$) gauge theory with an imaginary $\theta$ term can be
nonpertubatively formulated on the lattice by
\begin{eqnarray}
&&Z_L=\int [dA] \exp \left(  - S_L + \theta_L Q_L \right) ,\label{ZL}\\
&& S_L = - {\beta\over N} \sum_{x,\mu>\nu} {\rm Re} {\rm Tr}\, 
\Pi_{\mu\nu}(x),\qquad \beta={2N\over g_0^2},
\label{wilsonac} 
\end{eqnarray}
where $g_0$ is the bare coupling, 
$\Pi_{\mu\nu}$ is the standard plaquette operator, given by the
product of link variables along a $1\times 1$ plaquette of the lattice
\begin{equation}
\Pi_{\mu\nu}(x) = U_\mu(x)U_\nu(x+\mu+\nu)U^\dagger_\mu(x+\nu)
U^\dagger_\nu(x),
\label{plaquette}
\end{equation}
and the topological term 
\begin{equation}
Q_L\equiv \sum_x q_L(x)
\label{qlat}
\end{equation}
is constructed using the lattice operator~\cite{DFRV-81}
\begin{equation}
q_{L}(x) = - {1\over 2^4\times 32 \pi^2} \sum^{\pm
4}_{\mu\nu\rho\sigma=\pm 1} \epsilon_{\mu\nu\rho\sigma} {\rm Tr}
\left[ \Pi_{\mu\nu}\Pi_{\rho\sigma}\right].
\label{qL}
\end{equation}
Notice that this is not the only possible choice for $q_L$\,; the only
requirement 
is that it must have the correct continuum limit (\ref{topchden}) when
$a\to 0$ ($a$ is the lattice spacing). Our choice of the operator
$q_L(x)$ is also motivated by its relative simplicity in view of the
MC simulations.  In the continuum limit~\cite{CDP-88} $q_L(x)$, being
a local operator, behaves as
\begin{equation}
q_{L}(x)\longrightarrow a^4 Z_q q(x) + O(a^{6}),
\label{renorm}
\end{equation}
where $Z_q$ is a finite function of the bare coupling, going to one in
the limit $\beta\rightarrow \infty$.  Thus, we have the correspondence
\begin{equation}
\theta_i = Z_q \theta_L,
\label{zth}
\end{equation}
apart from $O(a^2)$ corrections.  The renormalization $Z_q$ may be
evaluated by MC simulation at $\theta=0$, computing
\begin{equation}
Z_q = { \langle Q Q_L \rangle_{\theta=0} \over \langle Q^2 \rangle_{\theta=0} 
}\,,
\label{zqc}
\end{equation}
where $Q$ is a topological estimator such as those obtained by the
overlap method~\cite{Neuberger-01,HLN-98} or the cooling
method~\cite{Teper-00}, which are not affected by lattice
renormalizations, and, more importantly, by nonphysical background
contributions arising when the correlations of 
topological charge densities are
measured at coincident points. Note that Eq.~(\ref{zqc}) assumes that
such contributions are also absent when we consider expectation values
of mixed products of $Q$ and $Q_L$; this hypothesis will be checked by
the scaling consistency of the results.  In particular, we expect that
the ratios
\begin{eqnarray}
&&{\langle Q \rangle_{\theta_i}\over \langle Q^2 \rangle_{\theta=0}}=
\theta_i  - 2 b_2 \theta_i^3 + 3 b_4 \theta_i^5 + ... ,\label{avq}\\
&& {\langle Q^2 \rangle^c_{\theta_i}\over \langle Q^2 \rangle_{\theta=0}}
= 1 - 6 b_2 \theta_i^2 + 15 b_4 \theta_i^4 + ... ,\label{avq2}
\end{eqnarray}
have a well defined continuum limit as functions of $\theta_i$\,.

We mention that similar approaches based on MC simulations with
imaginary $\theta$ terms have been also pursued to study the $\theta$
dependence of 2D CP$^{N-1}$ models, see
Refs.~\cite{ADGL-02,ADGL-04,AGL-07,AP-07}.

In the following we present results of MC simulations of the 4D SU(3)
lattice gauge theory, at $\beta=5.9,\,6,\,6.2$, for lattice sizes
$L=16,\,16,\,20$, respectively; the simulations are carried out both
at $\theta_L=0$ and $\theta_L\ne 0$, within the region
$|\theta_i|\lesssim\pi$.  The updating of the link variables is
performed using the overrelaxation algorithm described in
Ref.~\cite{DPRV-02}.  Since our numerical study requires
high-statistics MC simulations, we choose the cooling method as
estimator of the topological charge $Q$, and in particular the
implementation outlined in Ref.~\cite{DPV-02}.  The topological charge
has been measured on cooled configurations (by locally minimizing the
lattice action), using the twisted double plaquette operator
(\ref{qL}).  As is well known, the sum over the whole lattice of the
twisted double plaquette, measured on cooled configurations, takes on
values $Q_t \simeq k \alpha$, where $k$ is an integer and $\alpha
\lesssim 1$.  We determine the typical value of $\alpha$ by minimizing
the average deviation of the twisted double plaquette from integer
multiples of $\alpha$, and then assign to $Q$ the integer closest to
$Q_t/\alpha$.  The factor $\alpha$ turns out to depend on $\beta$, and
approaches one with increasing $\beta$; for example $\alpha\approx
0.93$ for $\beta=6$ and $\alpha\approx 0.95$ for $\beta=6.2$, at the
20$^{\rm th}$ cooling step.  This method eliminates the need for
expensive, protracted cooling; usually fewer than 20 steps suffice
(when using $N(N-1)/2$ subgroups) in order to observe a substantial
convergence of the results, which appears to improve with increasing
$\beta$.  The data that we will report are taken after 20 cooling
steps. This cooling method to estimate $Q$ is significantly less
expensive than the overlap method, although less
rigorous~\cite{VP-09}.

\hfill
\TABLE[ht]{
\caption{ 
$\theta=0$ MC results for the 4D SU(3) lattice gauge theory.
}
\label{tabtheta0}
\footnotesize
\begin{tabular}{lrllll}
\hline\hline
\multicolumn{1}{c}{$\beta$}&
\multicolumn{1}{c}{$L$}&
\multicolumn{1}{c}{$\chi\equiv \langle Q^2\rangle_{\theta=0}/V$}&
\multicolumn{1}{c}{$b_2$}&
\multicolumn{1}{c}{$b_4$}&
\multicolumn{1}{c}{$Z_q$}
\\
\hline\hline
5.9 & 16 & 0.0001532(4) & $-$0.026(6) &  $-$0.016(13) & 0.1122(6) \\ 

6.0 & 16 & 0.0000743(2) & $-$0.027(4) & $\phantom{-}$0.004(3) & 0.1353(9) \\ 

6.2 &  20 & 0.0000201(1)  & $-$0.028(4) &  $\phantom{-}$0.000(2) & 0.174(2)
\\ 

\hline\hline
\end{tabular}
}\hfill

\FIGURE[ht]{
\epsfig{file=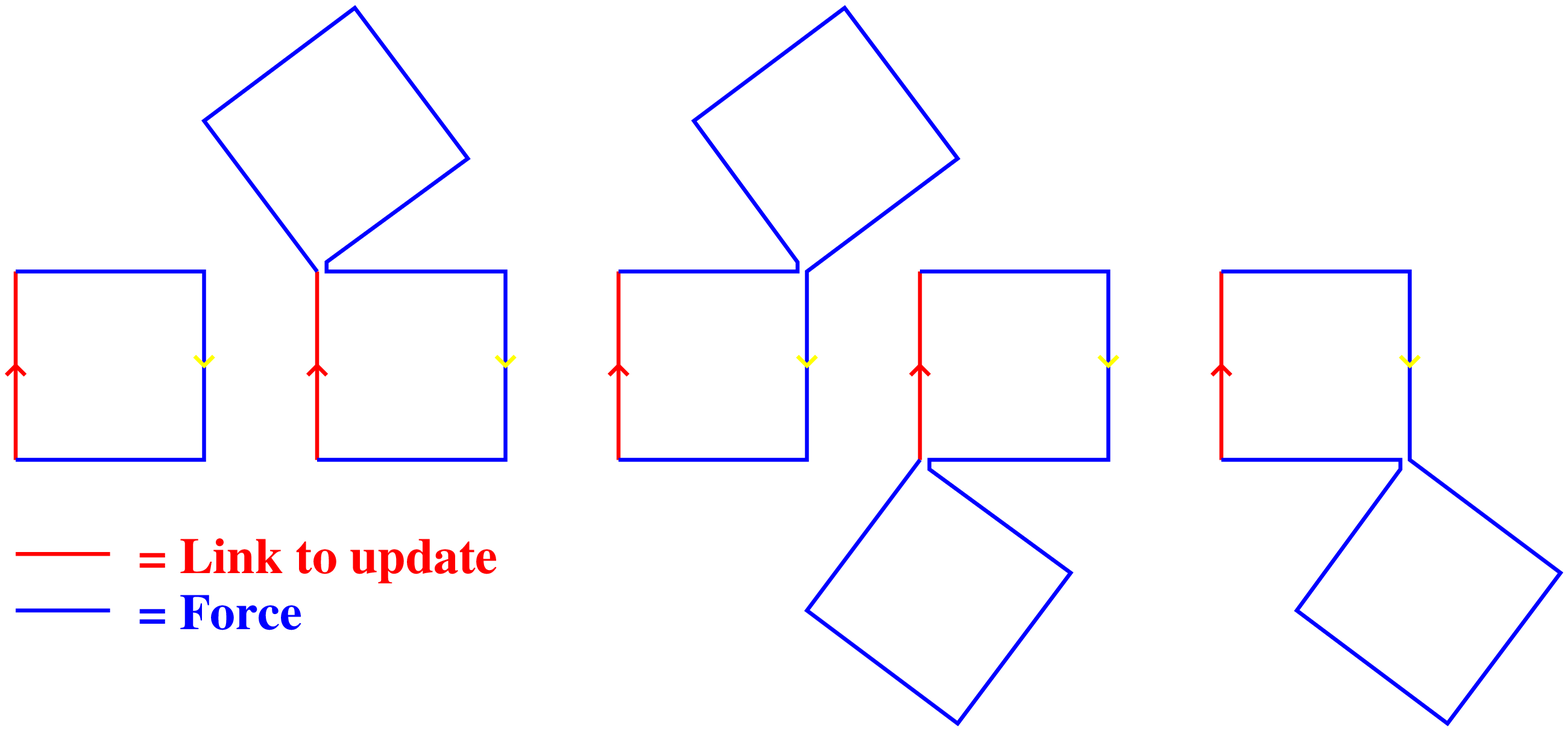, width=10truecm} 
\caption{The force corresponding to a given link. All but the first
  contribution arise from the $\theta$ term.
}
\label{staples}
}

To begin with, we report results from high-statistics MC simulations
at $\theta=0$. They required long MC runs,
essentially because of the severe (apparently exponential) critical
slowing down of the topological modes~\cite{DPV-02}, whose
autocorrelation time (in units of sweeps) ranges from
$\tau_Q \approx 77$ at $\beta=6$ to $\tau_Q\approx 540$ at
$\beta=6.2$. In particular, over 40 million sweeps per value of
$\beta$ were produced on average for runs at $\theta = 0$.  
The results for the topological susceptibility $\chi$,
the coefficients $b_2$ and $b_4$, and the renormalization
constant $Z_q$, are reported in Table~\ref{tabtheta0}.
The estimates of $\chi$ and $b_2$ are consistent with earlier MC
simulations, see, e.g., \cite{DPV-02}.  The estimate of high-order
coefficients, such as $b_4$, turns out to be very hard in $\theta=0$
MC simulations, requiring huge statistics, due to the large
cancellations in their expressions in terms of correlators at
$\theta=0$, see Eq~(\ref{b4chi6}), giving rise to relatively large
statistical errors.  The results for $b_4$ are consistent with zero,
suggesting the bound $|b_4|\lesssim 0.005$, which will be improved
below. The estimates of $Z_q$ may be compared with the one-loop
calculation~\cite{CDP-88} $Z_q = 1 - 5.448 \beta^{-1} +
O(\beta^{-2})$, which underestimates it, and the estimates by the
heating method at $\beta=6$: $Z_q=0.18(1)$~\cite{ACDDPV-94} and
$Z_q=0.16(1)$~\cite{Delia-03}, which are slightly larger than the
values that we obtain in our MC simulations using Eq.~(\ref{zqc}).

The MC simulations at $\theta_L\ne 0$ are quite slower, due to the
more complex structure of the action terms involving a single link.
The ``force'' corresponding to a given link now consists of additional
contributions, beyond the standard sum of staples, as shown in
Fig.~\ref{staples}. Thus, the link updating procedure is slower by
approximately a factor of three with respect to the $\theta_L=0$ case.
In runs with $\theta_L\ne 0$, an average of 3
million sweeps were produced for each value of $\beta$ and $\theta_L$.

\FIGURE[ht]{ \epsfig{file=distq6p2.eps, width=10truecm}
\caption{Distribution of the ratio $Q_t/\alpha$, for $\beta=6.2$
configurations at $\theta_L = 0$ and $\theta_L=8.7$
(corresponding to $\theta_i\approx 1.5$), after 20
cooling steps.  }
\label{histogram}
}

The cooling algorithm is implemented in the same way as for the
$\theta_L=0$ simulations. The presence of a $\theta_L$ term with
$\theta_L > 0 \ (\theta_L<0)$ leads to a preponderance of
configurations with $Q > 0 \ (Q < 0)$. Fig.~\ref{histogram} shows the
distributions of the ratio $Q_t/\alpha$ of $\beta=6.2$ cooled
configurations at $\theta_L = 0$ and $\theta_L = 8.7$
(corresponding to $\theta_i=Z_q\theta_L\approx 1.5$), after 20
cooling steps; we recall that the
topological estimator $Q$ is obtained by taking the integer closest to
$Q_t/\alpha$, $\alpha(\beta=6.2)=0.95$.
We note that these distributions cluster around integer
values, also for rather large values of $Q_t/\alpha$, both for $\theta_L=0$
and $\theta_L=8.7$, although the optimal value $\alpha=0.95$ is
kept fixed.  The autocorrelation time of the topological charge in the
MC simulations at $\theta_L\ne 0$ does not significantly change in
units of sweeps with respect to that at $\theta_L=0$ and the same
$\beta$.

\FIGURE[ht]{
\epsfig{file=qtheta.eps, width=10truecm} 
\caption{ The ratio $\langle Q \rangle_{\theta_L}/\langle Q^2
\rangle_{\theta=0}$ vs $\theta_i=Z_q\theta_L$.  Here scaling
corrections are only visible for $\theta_i>2$. The dashed line shows
the curve $\theta_i-2b_2\theta_i^3$ with $b_2=-0.026$, while the
dotted line shows the simple linear behavior.  }
\label{qtheta}
}

Figs.~\ref{qtheta} and \ref{rqoq2} show MC results for
the ratios $\langle Q \rangle_{\theta_L}/\langle Q^2
\rangle_{\theta=0}$ and $\langle Q \rangle_{\theta_L}/\langle Q^2
\rangle^c_{\theta_L}$, respectively, versus $\theta_i=Z_q\theta_L$.  The MC data at
different $\beta$ values appear to follow the same curve, providing
evidence of scaling.  Scaling corrections, which are expected to be
$O(a^2)$, appear quite small, and tend to increase with increasing
$\theta_i$.  This good scaling behavior thus corroborates the
existence of a nontrivial continuum limit for any value of $\theta_i$.

In Table~\ref{tabfits} we present the results of fits of the data for
the ratio $\langle Q \rangle_{\theta_L}/\langle Q^2
\rangle_{\theta=0}$, to the polynomial Ansatz
\begin{equation}
{\langle Q \rangle_{\theta_L}\over \langle Q^2 \rangle_{\theta=0}}=
Z_q \theta_L - 2 b_2 (Z_q\theta_L)^3 + 
3 b_4 (Z_q\theta_L)^5 + ...
\label{fitans}
\end{equation}
The results turn out to be quite stable with respect to the maximum
value of $|\theta_i|$ allowed in the fits.  They are in good agreement
with the $\theta=0$ results presented in Table~\ref{tabtheta0},
improving them significantly. In particular, they provide a much
smaller bound on the value of $b_4$.  The expected $O(a^2)$ scaling
corrections in the estimates of $b_{2i}$ are taken into account by
extrapolating the results at finite  $\beta$ fitting them to
\begin{equation}
b_{2i} + c \,\sigma(\beta)
\label{exb2}
\end{equation}
where $\sigma$ is the string
tension in unit of the lattice spacing (we use the data of
Ref.~\cite{DPV-02}).  This provides an accurate estimate of $b_2$,
that is
\begin{eqnarray}
b_2=-0.026(3),
\label{b2est}
\end{eqnarray}
whose error includes statistical and systematic errors related to the
extrapolation to the continuum limit and to the small differences of
the fit results reported in Table~\ref{tabfits}.  This estimate of
$b_2$  is
clearly more precise than the one we may obtain by a continuum-limit
extrapolation of the data at $\theta=0$ reported in
Table~\ref{tabtheta0} using Eq.~(\ref{exb2}), which gives
$b_2=-0.029(7)$.  Moreover, the results of Table~\ref{tabfits}
lead to  a very small bound for $b_4$:
\begin{eqnarray}
|b_4|< 0.001,
\label{b4est}
\end{eqnarray}
which is much smaller than the one obtained from the $\theta=0$
simulations.  These results confirm that the coefficients $b_{2i}$ of
the expansion (\ref{Fthetai}) are rapidly decreasing with increasing
$i$, as expected by the large-$N$ prediction
$b_{2i}=O(N^{-2i})$~\cite{VP-09}.

\FIGURE[ht]{
\epsfig{file=rqoq2.eps, width=10truecm} 
\caption{ The ratio $\langle Q \rangle_{\theta_L}/ \langle Q^2
\rangle^c_{\theta_L}$ vs $\theta_i=Z_q\theta_L$. }
\label{rqoq2}
}

The results (\ref{b2est}) and (\ref{b4est}) significantly improve
earlier results for the expansion of the ground-state energy around
$\theta=0$, as obtained by $\theta=0$ MC simulations using different
methods to estimate $Q$: $b_2=-0.023(7)$ from Ref.~\cite{DPV-02},
$b_2=-0.024(6)$ from Ref.~\cite{Delia-03}, and $b_2=-0.025(9)$ from
Ref.~\cite{GPT-07}, while no estimates for $b_4$ had been reported.

In conclusion, we have investigated the scaling behavior of the 4D
SU(3) gauge theory in the presence of an imaginary $\theta$ term, by
MC simulations computing the topological properties at imaginary
$\theta$, i.e.  $\theta=-i\theta_i$ with real $\theta_i$.  The
numerical results for the topological charge provide a good evidence
of scaling in the region $|\theta_i|\lesssim \pi$ which we consider.
The imaginary $\theta$ dependence of the ground-state energy turns out
to be well described by the first few nontrivial terms of the
expansion around $\theta=0$.  Fits to polynomials provide a quite
accurate estimate of $b_2$ and a very small bound on $b_4$, see
Eqs.~(\ref{b2est}) and (\ref{b4est}), which support the expected
large-$N$ scenario predicting $b_{2i}=O(N^{-2i})$.  This study may be
straightforwardly extended to other observables, to determine their
$\theta$ dependence, see Ref.~\cite{DMPSV-06}.  Finally, we mention
that, besides allowing more precise determinations of the $\theta=0$
expansion coefficients of the ground-state energy and other
observables, the use of imaginary $\theta$ values might turn out
useful in the effort to overcome the problem of the dramatic critical
slowing down of the topological
modes~\cite{VP-09,DPV-02,DMV-04,CRV-92,alpha-11}, by performing parallel
tempering simulations~\cite{par-temp} with a set of imaginary $\theta$ values
including $\theta=0$, which provides an {\em exact} MC algorithm for
the model at $\theta=0$. This approach is largely used in the MC
simulations of spin-glass models which are affected by analogous
critical slowing down problems.

\hfill
\TABLE[ht]{
\caption{ Results of some fits of the data of ${\langle Q
\rangle_{\theta_L}/\langle Q^2 \rangle_{\theta=0}}$ to $O(\theta_L^k)$
odd polynomials.  The fits use also the $\theta=0$ results of
Table \ref{tabtheta0} and consider only data up to a maximum value
$|\theta_i|_{\rm max}$ of $|\theta_i|$, leading to an acceptable
$\chi^2$, $\chi^2/{\rm dof}\lesssim 2$.  }
\label{tabfits}
\footnotesize
\begin{tabular}{lccllll}
\hline\hline
\multicolumn{1}{c}{$\beta$}&
\multicolumn{1}{c}{order of the polynomial}&
\multicolumn{1}{c}{$|\theta_i|_{\rm max}$}&
\multicolumn{1}{c}{$Z_q$}&
\multicolumn{1}{c}{$b_2$}&
\multicolumn{1}{c}{$b_4$}\\
\hline\hline
5.9 & $O(\theta_L^3)$ & 1.0 & 0.1121(5)  & $-$0.030(3) & \\
  &   &  1.5 & 0.1122(4) & $-$0.0285(14) & \\
  &   &  2.0 & 0.1122(3) & $-$0.0287(8) & \\ 
  &   $O(\theta_L^5)$ & 1.0 & 0.1120(5)  & $-$0.025(6) & 0.004(4) \\
  &   &  1.5 & 0.1122(4) & $-$0.029(4) & 0.000(1) \\
  &   &  2.0 & 0.1122(4) & $-$0.029(2) & 0.0000(3) \\\hline

6 &  $O(\theta_L^3)$ & 1.0 & 0.1346(4)  & $-$0.028(2) & \\
  &   &  1.3 & 0.1346(3) & $-$0.0285(14) & \\
  &   &  1.7 & 0.1346(3) & $-$0.0282(9) & \\
  &   &   2.0 & 0.1346(3) & $-$0.0282(6) & \\  
 &  $O(\theta_L^5)$ & 1.0 & 0.1345(4)  & $-$0.027(4) & 0.002(3) \\
  &    &  1.3 & 0.1346(4) & $-$0.028(3) & 0.000(1) \\
  &    &  1.7 & 0.1346(4) & $-$0.028(2) & 0.0000(5)\\
  &    &   2.0 & 0.1346(4) & $-$0.028(2) & 0.0000(2) \\\hline

 6.2 &  $O(\theta_L^3)$ & 1.0 & 0.1717(13)  & $-$0.027(4) & \\
  &   &  1.5 & 0.1716(13) & $-$0.027(3) & \\
  &   &  2.0 & 0.1717(13) & $-$0.027(2) & \\
  &  &  2.7  &  0.1717(12) & $-$0.0265(13) & \\ 
 &   $O(\theta_L^5)$ & 1.0 & 0.1724(15)  & $-$0.027(4) & $-$0.002(6) \\
  &   &  1.5 & 0.1722(13) & $-$0.027(4) &   $-$0.0003(14) \\
  &   &   2.0 & 0.1722(13) & $-$0.027(3) & $-$0.0002(6) \\
  &   &  2.7 & 0.1721(12) & $-$0.026(3)  & $-$0.0000(2) \\

\hline\hline
\end{tabular}
}\hfill

\end{document}